# Geography as a Science of the Earth's Surface Founded on the Third View of Space

Bin Jiang

Faculty of Engineering and Sustainable Development, Division of GIScience
University of Gävle, SE-801 76 Gävle, Sweden
Email: bin.jiang@hig.se

*(Draft: August 2020, Revision: January, March, July, August 2021)*

**Abstract:**
The third (or organismic) view of space states that space is neither lifeless nor neutral, but a living structure capable of being more living or less living, thus different fundamentally from the first two mechanistic views of space: Newtonian absolute space and Leibnizian relational space. The living structure is defined as a physical and mathematical structure or simply characterized by the recurring notion (or inherent hierarchy) of far more small substructures than large ones. This paper seeks to lay out a new geography as a science of the Earth's surface founded on the third view of space. The new geography aims not only to better understand geographic forms and processes but also – maybe more importantly – to make geographic space or the Earth's surface to be living or more living. After introducing two fundamental laws of geography: Tobler's law on spatial dependence (or homogeneity) and scaling law on spatial heterogeneity, we argue that these two laws are fundamental laws of living structure that favor statistics over exactitude, because the former (or statistics) tends to make a structure more living than the latter (or exactitude). We present the concept of living structure through some working examples and make it clear how a living structure differs from a non-living structure, under the organismic worldview that was first conceived by the British philosopher Alfred Whitehead (1861–1947). In order to make a structure or space living or more living, we introduce two design principles – differentiation and adaptation – using two paintings and two city plans. The new geography is a science of living structure, dealing with a wide range of scales, from the smallest scale of ornaments on walls to the scale of the entire Earth's surface.

## 1. Tobler's law and scaling law of geography

As charmingly stated by Tobler (1970), *"everything is related to everything else, but near things are more related than distant things"*. This is known as Tobler's law or the first law of geography, implying that space and time are not random but auto-correlated. Your housing price is more closely related to those of your neighbors than to those of your neighbors' neighbors. Today's weather is more related to yesterday's than to the day before yesterday's. Two locations that are one meter away are more related than two locations that are 10 meters away. Tobler's law is commonly seen not only in space and time, but also in society. You are more related to your friends than to the friends of your friends; you are more genetically related to your parents than to the parents of your parents. Clearly, Tobler's law is not just limited to geography, but applies to social sciences, biology and many others.

Tobler's law has two underlying keywords: relatedness and nearness. The notion of relatedness refers to how things are similar or dissimilar to each other. What Tobler's law expresses essentially is a kind of similarity (or dissimilarity) or homogeneity, which can be characterized by an average or mean under Gaussian thinking (Jiang 2015). We can fairly predict your housing price by averaging those of your neighbors' houses. Today is very likely to be a sunny day because yesterday was a sunny day. The keyword nearness indicates that things are related to or similar (or dissimilar) to each other at local or nearby scales rather than a global scale. In other words, Tobler's law holds on each scale (rather than across scales) reflecting more or less similar things nearby: the nearer two things are, the more related (similar or dissimilar) they are; or conversely, the more distant they are, the less related (similar or dissimilar) they are. Related to these two keywords is the fact that topological connection makes better sense than geometric details, as reflected in central place theory (Christaller 1933, 1966, Jiang 2018); see more detail in Figure 1 and related discussions.

In contrast to more or less similar things on each scale, there are far more small things than large ones across scales ranging from the smallest to the largest. The notion of far more smalls than larges was formulated as the scaling law (Jiang 2015). More or less similar things occurs (or recurs, to be more precise) on each scale, while far more small things than large ones recurs across scales or on a global scale. The notion of far more smalls than larges, which is also called spatial heterogeneity or spatial hierarchy, is typical of many societal and natural systems (Bak 1996, Simon 1996). There are far more low housing prices than high ones in a city; there are far more ordinary weather conditions than extraordinary ones over time; there are far more ordinary people than extraordinary people in any country or society. Thus, the scaling law is not just limited to geography, but applies universally to many other sciences.

There are four points to note about the scaling law. First, the scaling law is not about more smalls than larges, but far more smalls than larges, with "far" indicating the disproportion between smalls and larges, and their occurring numbers. Second, the notion of far more smalls than larges recurs multiple times rather than occurs just once, hence the recurring notion of far more smalls than larges. Third, the recurring notion of far more smalls than larges at different levels of scale (or across scales) is related to each other to form an inherent, coherent hierarchy. In other words, a very few largest things, numerous smallest things, and some in between the largest and the smallest constitute a coherent whole. This expression of the scaling law resembles the laws of architecture (Salingaros 1995), but these laws assume that the things' sizes strictly follow a power law distribution. Fourth, the scaling law does not require any strict mathematical distributions such as a power law and lognormal. Instead, it simply relies on head/tail breaks (Jiang 2013; see Section 3 for a working example) to derive the inherent hierarchy, which is the number of times the notion of far more smalls than larges recurs. The scaling law holds if the notion of far more smalls than larges recurs at least twice.

This paper is attempted to setup geography on these two laws and under the third or organismic view of space: space is neither lifeless nor neutral, but a living structure capable of being more living or less living (Alexander 2002–2005, 1999). Living structure is such a structure that consists of far more small substructures than large ones across all scales ranging from the smallest to the largest (the scaling law), yet with more or less similar sized substructures on each of the scales (Tobler's law). Therefore, living structure is said to be governed by these two fundamental laws (Jiang 2019). Among the two laws, the scaling law is the first, or dominant law, as it is universal, global, and across scales, while Tobler's law is available locally or on each of the scales. Conventionally, geography has been viewed as a minor science or an applied science that seeks to use or apply major sciences for understanding geographic forms and processes (c.f., Note 1 for more details). In this paper, we argue that the new geography is a major science, a science of living structure, not only for understanding geographic forms and processes, but also for making and remaking geographic space or the Earth's surface towards a living or more living structure.

The remainder of this paper is organized as follows. Section 2 argues that the two laws of geography are fundamental laws of living structure that favors statistics over exactitude. Section 3 presents some examples to differentiate living structure from non-living one under two different world views. Section 4 briefly introduces the two world views: Cartesian mechanistic and Whitehead's organismic. Section 5 illustrates two design principles differentiation and adaptation in order to make or transform a space to be living or more living. Section 6 further discusses the new geography and its deep implications. Finally, the paper concludes with a summary pointing to a prosperous future of the new geography.

## 2. Two laws together for characterizing living structure

Tobler's law and the scaling law are not only two laws of geography, but also – more importantly – fundamental laws of living structure. The notion of living structure applies to all organic and inorganic phenomena in the scales ranging from the smallest Planck's length to the largest scale of the universe (Alexander 2002–2005, 2003), so do these two laws. The applicability implies that there are far more small particles than large ones, far more rats than elephants, far more small stars than large ones, far more small galaxies than large ones, and so on. This paper deals with a range of scales of the Earth's surface between $10^{-2}$ and $10^6$ meters. Table 1 shows how these two laws complement to each other from various perspectives (Jiang and Slocum 2020). It is wise to keep the scaling law as the dominant one, as it is global or across scales, whereas Tobler's law is local or on each scale. However, in conventional geography, Tobler's law is usually overstated as the first law of geography, and it implies that the Earth's surface is in a simple and well-balanced equilibrium state. However, we know that the Earth's surface is unbalanced and very heterogeneous and every place is unique (Goodchild 2004). Dominated by the scaling law or the non-equilibrium character, the new geography aims not only to better understand the complexity of the Earth's surface, but also to make the Earth's surface a living or more living structure. For creating living structures, two design principles – differentiation and adaptation – will be introduced later on.

Table 1: Two complementary laws of geography or living structure

| Scaling law | Tobler's law |
| --- | --- |
| There are far more small things than large ones | There are more or less similar things |
| across all scales, and | available at each scale, and |
| the ratio of smalls to larges is disproportional (80/20). | the ratio of smalls to larges is closer to proportional (50/50). |
| Globally, there is no characteristic scale, so exhibiting | Locally, there is a characteristic scale, so exhibiting a |
| Pareto distribution, or a heavy-tailed distribution, | Gauss-like distribution, |
| due to spatial heterogeneity or hierarchy, indicating | due to spatial homogeneity or dependence, indicating |
| complex and non-equilibrium character. | simple and equilibrium character. |

Unlike many other laws in science, these two laws are statistical rather than exact. The statistical nature is more powerful than the exactitude one. Below, we cite three sets of evidence in science and art to make it clear why exactitude is less important. First, Zipf's law (1949) is also statistical rather than exact. It states that in terms of city sizes, the largest city is about twice as big as the second largest, approximately three times as big as the third largest, and so on. Here twice, three times, and so on are not exact, but statistical or roughly. Among the two sets: [1, 1/2, 1/3, …, 1/10] and [1 + $e_1$, 1/2 + $e_2$, 1/3 + $e_3$, …, 1/10 + $e_{10}$] (where $e_1$, $e_2$, $e_3$, … $e_{10}$ are very small values), the first dataset does not follow Zipf's law, while the second does. Zipf's law is a major source of inspirations of fractal geometry (Mandelbrot 1982). In his autobiography, Mandelbrot (2012) made the following remark while describing the first time he was introduced to a book review on Zipf's law: *"I became hooked: first deeply mystified, next totally incredulous, and then hopelessly smitten … to this day. I saw right away that, as stated, Zipf's formula could not conceivably be exact."* A dataset following Zipf's law meets the scaling law, but not vice versa, which means that the scaling law is even more statistical than Zipf's law. Zipf's law requires a power law while the scaling law does not.

The second evidence is not only statistical, but also geometrical. The leaf vein shown in Figure 1 (Jiang and Huang 2021) apparently has far more small substructures than large ones from the largest square to the smallest white spots. Carefully examining the structure of the leaf vein, it is not difficult to find that there are four different levels of scale according to thickness of their outlines. In contrast, the Sierpinski carpet also has far more smalls than larges; that is, far more small squares than large ones, exactly rather than statistically (Sierpinski 1915). Let us carefully examine the exactitude of the carpet. The largest square in the middle of the carpet is size 1/3, which is surrounded by eight squares of size 1/9, each of which is surrounded by eight squares of size 1/27, each of which is surrounded by eight squares of size 1/81. Thus, there are two exponential data series, each of which is controlled by some exact number. The size of squares is exponentially decreased by the exact number 1/3: (1/3, 1/9, 1/27, 1/81), whereas the number of squares is exponentially increased by the exact number 8: (1, 8, 64, 512). Clearly there are far more small squares than large ones exactly rather than statistically. Because of the exactitude, the Sierpinski carpet is less living structurally than the leaf vein. The first two sets of evidence are apparently from science, while the third set is from art.

The French painter Henri Matisse (1947) made a famous statement about the essence of art: *"Exactitude is not truth"*. In terms of exactitude, a photo is far better than a painting. However, the value of a painting lies not in its exactitude, but in something else, which is not only inexact, but also distorted or exaggerated. The distorted or exaggerated nature is often used in drawing a cartoon. A human face is a living structure governed by the scaling law and Tobler's law, with the recurring notion of far more smalls than larges. The eyes, nose, mouth, and ears are the largest features and are therefore the most salient; each of them – if examined carefully – is a living structure again, with the recurring notion of far more smalls than larges. All human faces are universally beautiful in terms of the underlying living structure, despite some tiny cultural effects on their beauty.

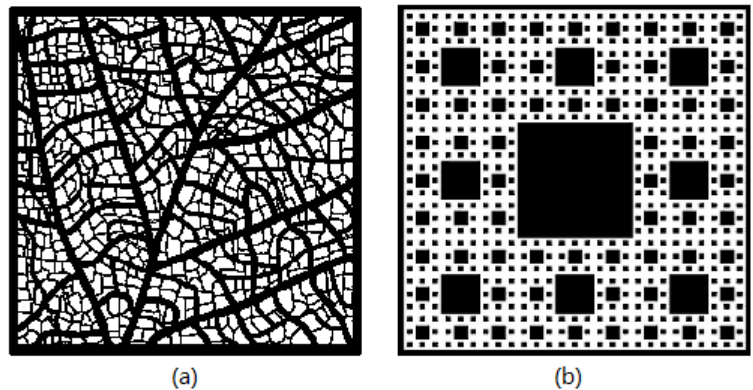

Figure 1: The leaf vein looks more living or more beautiful than the stiff Sierpinski carpet
(Note: The leaf vein (a) and the Sierpinski carpet (b) both meet the scaling law and Tobler's law, but the leif vein is more living than the Sierpinski carpet. This is because the squares of the Sierpinski carpet on each scale are precisely the same rather than more or less similar, thus violating Tobler's law to some extent. Another obvious reason that the Sierpinski carpet is less beautiful is that its negative space (white part) is not well shaped (based on one of the 15 properties, see Table 2), although its positive space is well shaped, by which we meant shapes are convex rather than concave. In contrast, both positive and negative spaces of the leaf vein are well shaped.)

The scaling law and Tobler's law are really two fundamental laws about livingness or beauty. They can be used to examine many patterns or structures (e.g., Wade 2006, Wichmann and Wade 2017) for understanding not only why they are beautiful, but also how beautiful they are. For example, the leaf vein is living or beautiful because of the recurring notion of far more small structures than large ones. This way, through these two laws, the livingness or beauty of a structure or pattern can be objectively or structurally judged. Importantly, the livingness judged through these two laws can be well reflected in the human mind and heart, thus evoking a sense of beauty. This point will be further discussed in the following.

**3. Living versus nonliving structure: The "things" the two laws refer to**
The two laws have a common keyword – "things": (1) more or less similar things on each scale, and (2) far more small things than large ones across all scales. What are the "things" the two laws refer to? We have provided examples above while introducing these two laws. For example, cities in a country and streets in a city are the things, for they are with far more smalls than larges. Seen from the perspective of cities and streets, the country and the city are living structures. In general terms, the things that collectively constitute a living structure are the right things, whereas the things that collectively do not constitute a living structure are not the right things. For example, if the leaf vein was saved as a gray-scale image with 1024 by 1024 pixels, each of which has a gray scale between 0 and 255, careful examination of these pixel values would show that they do not have far more light (or dark) pixels than dark (or light) ones. This way, we would end up with an absurd conclusion that the leaf vein is not a living structure. In fact, the pixels are not the right things, or the pixel perspective is not the right perspective for seeing the living structure.

In addition to the perspective discussed above, the scope also matters in seeing a living structure. A tree has surely far more small branches than large ones across scales from the largest to the smallest, while branches on each scale are more or less similar. Thus, the tree is no doubt a living structure, not biologically but in terms of the underlying structure. However, its leaves can be both living and non-living structure depending on the scope we see them. If we go down to the scope or scale of intra-leaves, each of them has multiple scales, showing a living structure (as shown in Figure 1). If we on the other hand concentrate on inter-leaves, they are all more or less similar sized, they are just the smallest scale of the tree. The leaf vein shown in Figure 1 is not a complete leaf, but part of it, with the large enough scope for us to see the living structure. All geographic features are living structures, if they are seen correctly with the right perspective and scope.

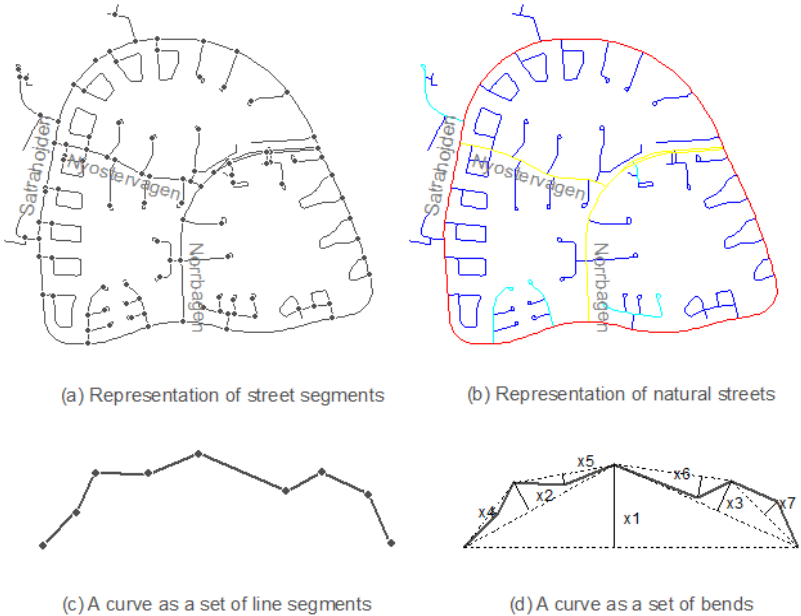

(a) Representation of street segments  (b) Representation of natural streets

(c) A curve as a set of line segments  (d) A curve as a set of bends

Figure 2: (Color online) Non-living versus living structure views of geographic features
(Note: Conventionally, a street network is represented as a set of geometric primitives, which are not the right things or substructures (a), whereas it is more correctly perceived as a collection of named streets, which are the right things or substructures for seeing the street network as a living structure (b). Each street is colored as one of the four levels of scale: blue for the least connected streets, red for the most connected street (only one), and yellow and turquoise for those between the most and the least connected. A coastline is conventionally represented as a set of line segments, which are not the right things or substructures (c), but it is more correctly perceived as a collection of far more small bends than large ones, which are the right things or substructures for seeing the coastline as a living structure (d). It is because the notion of far more small bends than large ones occurs twice: (1) $x_1 + x_2 + x_3 > x_4 + x_5 + x_6 + x_7$, and (2) $x_1 > x_2 + x_3$.)

Let us further clarify the term "things" through two working examples: A street network and a coastline (Figure 2, Jiang and Slocum 2020). Conventionally, in geography or geographic information science, the things often refer to geometric primitives such as pixels, points, lines, and polygons. It is little wonder that Tobler's law is seen pervasively, as there are more or less similar sized things seen from the perspective of geometric primitives. For example, a street network has more or less similar street segments, or all the street junctions have more or less similar numbers of connections (1–4) (Figure 2a). A coastline consists of a set of more or less similar line segments (Figure 2c). Unfortunately, all these geometric primitives are not the right things for seeing the street network or coastline as a living structure. There is little wonder, constrained by the geometric primitives, that living structure was not a formal concept in geography.

A street network is more correctly conceived of as a set of far more short streets than long ones, or a set of far more less connected streets than well connected ones (Figure 2b). The street network has four levels of scale, indicated by the four colors, far more short streets than long ones across the scales, and more or less similar streets on each of the four scales. A coastline is more correctly represented as a set of far more small bends than large ones (Figure 2d). The coastline has three levels of scale, indicated by three sets of bends: [$x_1$], [$x_2$, $x_3$], and [$x_4$, $x_5$, $x_6$, $x_7$]. The notion – or the recurring notion – of far more smalls than larges should be the major criteria for whether things are the right things that enable us to see a living structure, or whether we have the right perspective and scope for seeing a living structure. As another example, within a large enough scope of time or space, there are far more ordinary weather conditions than extraordinary ones, whereas within a limited scope of time (10 days) or space (a city), weather conditions may be more or less similar.

The "things" that collectively constitute a living structure are also called centers (Alexander 2002–2005), a term that was initially inspired by the notion of organisms conceived by Whitehead (1929). Centers or organisms are the building blocks of a living structure, and their definitions are somewhat obscure. Instead, in this paper we use substructures to refer to the right things for seeing a living structure. This way, a living structure can be stated – in a recursive manner – as the structure of the structure of the structure, and so on. The things or substructures constitute an iterative system. To make the point clear, it is necessary to introduce the head/tail breaks (Jiang 2013), a classification scheme for data with a heavy-tailed distribution.

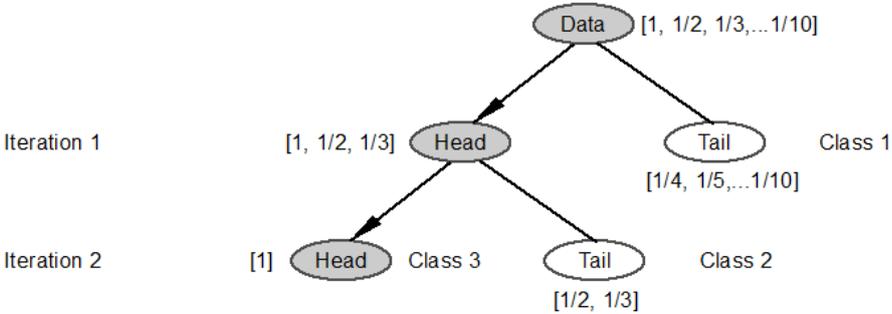

Figure 3: Head/tail breaks with a simple example of the 10 numbers.
(Note: The 10 numbers [1, 1/2, 1/3, …, 1/10] are classified into three classes: [1/4, 1/5, …, 1/10], [1/2, 1/3], and [1], which can be said to have three inherent hierarchical levels. The dataset, due to its inherent hierarchy, is therefore more living or more structurally beautiful than another dataset [1, 2, 3, …, 10], which lacks any inherent hierarchy, or violates the scaling law.)

For the sake of simplicity, we use the 10 numbers [1, 1/2, 1/3, …, 1/10] to show how they are classified through the head/tail breaks (Figure 3, Jiang and Slocum 2020). The dataset is a whole, and its average is about 0.29, which partitions the whole into two subwholes: those greater than the average are called the head [1, 1/2, 1/3], and those less than the average are called the tail [1/4, …, 1/10]. The average of the head subwhole is about 0.61, and it partitions the head subwhole into two subwholes again: those greater than the average are called the head [1], and those less than the average are called the tail [1/2, 1/3]. Instead of expressing the dataset as a set of numbers, we state the 10 numbers as an iterative system consisting of three subwholes recursively defined: [1], [1, 1/2, 1/3], and [1, 1/2, 1/3, …, 1/10]. Instead of perceiving these numbers as a set of 10 numbers, we consider them as a coherent whole, consisting of three subwholes including the whole itself. Or alternatively, these numbers as a coherent structure consists of three substructures including the structure itself. The dataset [1, 1/2, 1/3, …, 1/10], because of its inherent hierarchy of 3, is more living than the other dataset [1, 2, 3, …, 10] that is without any inherent hierarchy, or violates the notion of far more smalls than larges.

Now let us apply the recursive way of stating a whole or structure into the street network illustrated in Figure 2. Seen from above, the sample street network consists of 50 streets at four hierarchical levels indicated by the four colors: red (r), yellow (y), turquoise (t) and blue (b). Instead of stating the street network as a set or as four classes, we state it as an iterative system consisting of four subwholes or substructures that are recursively defined: [r], [r, $y_1$, $y_2$], [r, $y_1$, $y_2$, $t_1$, $t_2$, $t_3$, $t_4$, $t_5$], and [r, $y_1$, $y_2$, $t_1$, $t_2$, $t_3$, $t_4$, $t_5$, $b_1$, $b_2$, $b_3$, …, $b_{42}$]. In the same way, it is not difficult to figure out the three recursively defined subwholes for the coastline: [$x_1$], [$x_1$, $x_2$, $x_3$], and [$x_1$, $x_2$, $x_3$, …, $x_7$]. This living structure representation is recursive and holistic, so it differs fundamentally from existing representations that tend to focus on segmented individuals or mechanistic pieces. An advantage of the living structure representation is that the inherent hierarchy of space is obvious. To this point, we have seen clearly how the right things constitute an iterative system, being a living structure consisting of far more smalls than larges.

## 4. Two distinct world views: Cartesian mechanistic and Whitehead's organismic

The above examples have shown the two distinct ways of seeing space or our world in general: nonliving structure and living structure. Under the nonliving structure view, things are fragmented pieces, even though they may be connected externally rather than internally to each other as a collection of the pieces. Under the living structure view, things are not individual or independent, but part of a coherently and internally connected whole; the same holistic world view had been advocated by the eminent quantum physicist and philosopher David Bohm (1980). The nonliving structure view is commonly held under the Cartesian world picture (Descartes 1637, 1954, Figure 4), which assumes something as a machine to understand how it works. It is a powerful model about our world, for what we have achieved in science over the past hundred years is largely attributed to the mental model. The mechanistic world view is so dominated in our thinking as if it were the only mental model, or even worse it may be considered to the world itself. The mechanistic mental model is limited when comes to design or creation, as it adopts the nonliving structure view under which goodness of designed or created things is sidelined as an opinion or personal preference rather than a matter of fact (Alexander 2002–2005). Equally, under the nonliving structure view – or, more specifically, under Newton's absolute and Leibniz's relational views of space – a geographic space is represented as a collection of geometric primitives, which tend to be *"cold and dry"* (Mandelbrot 1982). With these nonliving structure representations (Figure 2a, 2c), one can hardly judge objectively how good a space or spatial configuration is.

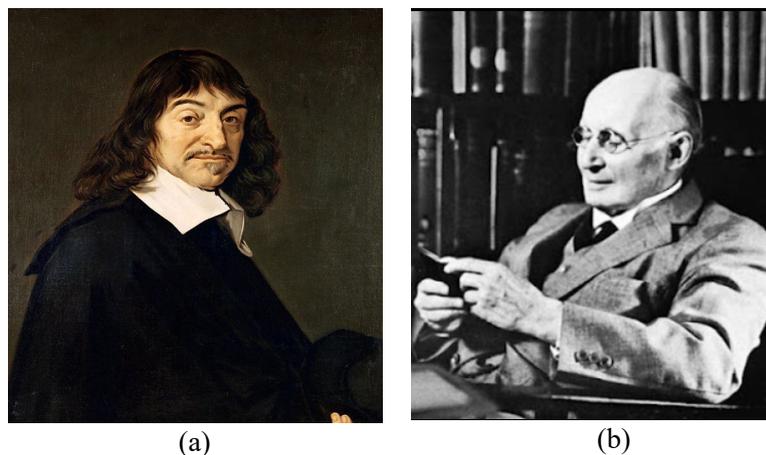

(a)           (b)

Figure 4: Two representative figures for the mechanistic and organismic world views
(Note: The French philosopher and mathematician René Descartes (1596–1650) who was commonly recognized as one of the scientists who formulated the mechanistic world view (a), and the English mathematician and philosopher Alfred North Whitehead (1861–1947) who conceived the organismic world view (b))

The Cartesian world picture, or the two views of space that are framed under the Cartesian cosmology, has two unexpected outcomes (Alexander 2002–2005). The first was that the "I" went out of the world picture and the inner experience of being a person is not part of this picture. The second was that the mechanistic world picture no longer has any definite feeling of value in it, or value has become sidelined as a matter of opinion rather than as something intrinsic to the nature of the world. The organismic world picture first conceived by Whitehead (1929) extends the mechanistic world picture to include human beings as part of the organismic world picture (Figure 4). The same world view has been promoted by David Bohm (1980) among many others. Under the organismic world view, we human beings are part of the world rather than separated from the world. Under the organismic mental model, a third view of space was conceived by Alexander (2002–2005): space is neither lifeless nor neutral, but a living structure capable of being more living or less living. Value lies on the underlying configuration of space and, in the end, goodness of space is no longer a matter of opinion, but a matter of fact. The shift from the opinion view to the fact view or from the mechanistic world view to the organismic world view represents something fundamental (Kuhn 1970) in our thinking about geography, for design or how to make living or more living space is at the forefront of geographic inquiry. In the following section, we will make it clear how space should become more living through two design principles.

**5. Two design principles: differentiation and adaptation**
In line with the two laws of living structure, there are two design principles – differentiation and adaptation – for transforming a space or structure to be living or more living. The purpose of the differentiation principle is to create far more small substructures than large ones, while the adaptation principle ensures that the created substructures are well adapted to each other, i.e., nearby substructures are more or less similar. These two design principles ensure that any geographic space would become living or more living from the current status. Importantly, goodness of a geographic space is considered as a fact rather than an opinion, as mentioned above. These two design principles are what underlie the 15 structural properties (Table 2) distilled by Alexander (2002–2005) from traditional buildings, cities and artifacts. The 15 structural properties can be used to transform a space or structure into living or more living structure. Interested readers should refer to Alexander (2002–2005), specifically Volumes 2 and 3, for numerous examples. In this section, we use two working examples – two paintings and two city plans – to clarify these two design principles.

Table 2: The 15 properties of living structure or wholeness

| Levels of scale | Good shape | Roughness |
|---|---|---|
| Strong centers | Local symmetries | Echoes |
| Thick boundaries | Deep interlock and ambiguity | The void |
| Alternative repetition | Contrast | Simplicity and inner calm |
| Positive space | Gradients | Not separateness |

The two paintings are not very living, as they meet only the minimum condition with three or four inherent hierarchical levels. Painting (a) by Dutch painter Piet Mondrian (1872–1944) is entitled *Composition II*, with the three colors of red, blue, and yellow, whereas painting (b) is modified slightly from painting (a) by the author (Jiang and Huang 2021). Figure 5 demonstrates how these two paintings are evolved – in a step-by-step fashion – from an empty square. Structurally speaking, painting (b) is more living than painting (a). It can equally be said that structure (g) is more living than structure (f), which is more living than (e), which is more living than structure (d), which is more living than structure (c). Thus, among all these structures or substructures, the empty square is the deadest, while (g) is the most living. On the one hand, there is the recurring notion of far more newborn substructures than old ones; on the other hand, within each iteration there are far more small substructures than large ones. Seen from the comparison, it is not hard to understand that one structure is – objectively – more living than another. As a reminder, this judgment is purely based on the underlying structure, having little to do with colors.

The goodness or livingness of a space – or a city in particular – is a matter of fact rather than an opinion or personal preference, based on the living structure. More specifically, the goodness of a space depends on substructures within the space, as we have already seen in the above discussion. The goodness also depends on larger space that contains the particular space; its context, so to speak. This way of judging goodness or order of things is universal across all cultures, faiths, and ethnicities, not only for natural things, but also for what we make or build. This is probably the single most important message in the masterful work *The Nature of Order* (Alexander 2002–2005, 2003). This is a radical departure from the current view of space in terms of its goodness, judged by various technical parameters such as density, accessibility, and greenness. The living structure constitutes the foundation of the new geography this paper seeks to advocate and promote.

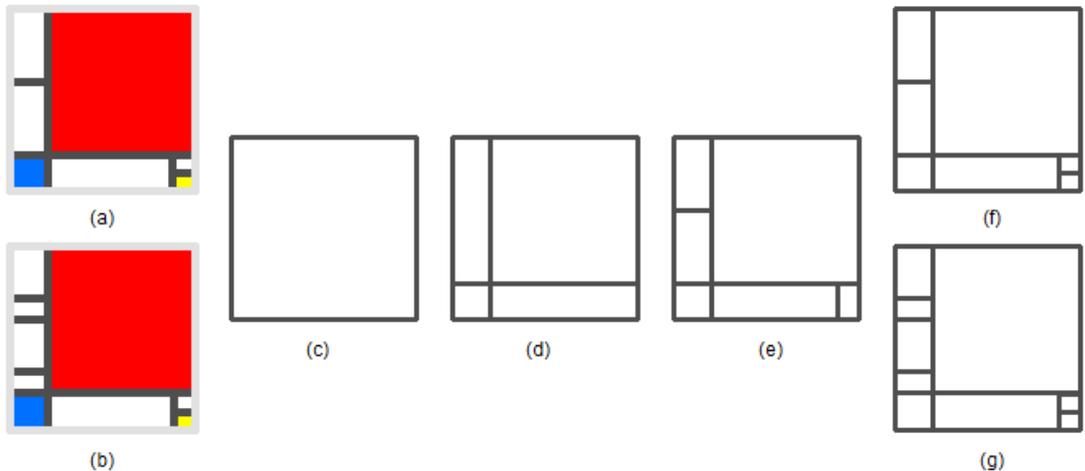

Figure 5: (Color online) Living and less-living structures and their differentiation processes. (Note: The two paintings – *Composition* (a) by the Dutch painter Piet Mondrian (1872–1944) and *Configuration* (b) modified from *Composition* by the author of this paper – meet the minimum condition of being a living structure. Both paintings can be viewed to be differentiated like cell division from the empty square (c), so they are featured by the recurring notion of far more newborn (newly generated) substructures than old ones. More specifically, there are far more newborns than old one from (c) to (d), and again from (d) to (e), except from (e) to (f) in which there is a violation of far more newborns than old ones. However, there is again far more newborns than old one from (e) to (g). On the other hand, in each iteration there are far more small substructures than large ones. Thus, the painting *Configuration* is more living or more beautiful – structurally – than the painting *Composition*. If the reader prefers *Composition* over *Configuration*, do not be panic and your preference is likely to be dominated by nonstructural factors such as cultures, faiths, and ethnicities. However, the kind of beauty determined by the underlying living structure accounts for the feelings shared by most people or peoples.)

The living structure perspective implies that a geographic space is in a constant evolution from less living to living or more living. Importantly, a geographic space or its design and planning process is an embryo-like evolution rather than LEGO-like assembly of prefabricated elements (Jiang and Huang 2021). Note that the evolution view differs fundamentally from the assembly view, with the former being organic or natural, while the latter being mechanical or less natural. The living structure perspective implies also that a structure or substructures must be seen recursively. For example, conventionally painting (a) is seen as composed of 7 pieces, but it is more correct to say it consists of 18 (1 + 4 + 6 + 7) recursively defined structures or substructures (Figure 5). Instead of being 9 pieces for painting (b), it is more correct to say that it consists of 20 (1 + 4 + 6 + 9) recursively defined structures.

Using the recursive perspective, it is not hard to understand why traditional city plans are usually more living than modernist counterparts. For example, with the city of London plan, the notion of far more small substructures than large ones recurs five times, so there are 6 hierarchical levels, whereas for the Manhattan one, the notion of far more small substructures than large ones recurs twice, so there are only 3 hierarchical levels (Figure 6). Thus, the left city plan is more living – structurally or objectively – than the right one. In the same fashion, traditional building facades are usually more living than modernist counterparts. In this regard, there have been many human perception tests supporting this conclusion (e.g., Alexander 2002–2005, Wu 2015), indicating over 75% agreement between the human perception and the reasoning based on the two laws. There may be some people (fewer than 25%) who prefer modernist buildings because they look new and luminous or for whatever personal reasons. A recent biometric investigation (Salingaros and Sussman 2020) has provided further evidence that traditional façades are more "engaging" with people than contemporary façades.

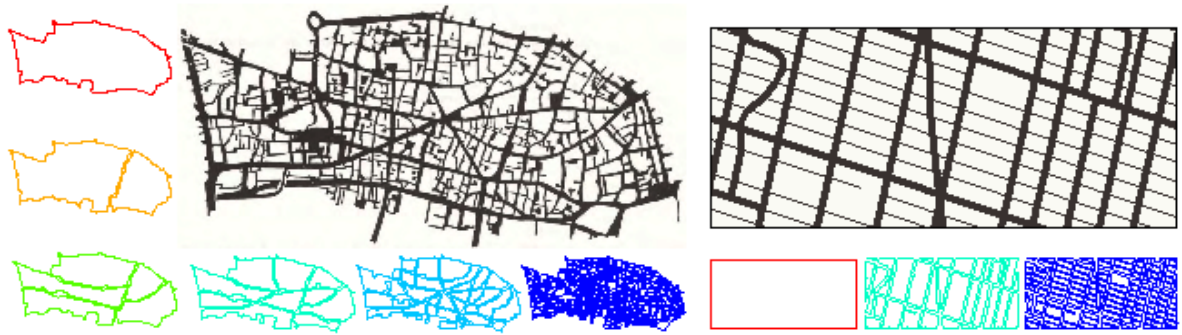

Figure 6: (Color online) Why the city of London plan is more living than the Manhattan one
(Note: The city of London plan (the left) is obviously a living structure, for it meets scaling law, or the recurring notion of far more small substructures than large ones across the 6 hierarchical levels, shown in colors in those reduced panels to the left (Jiang and Huang 2021). The part of Manhattan plan (the right) is less living, with only 3 inherent hierarchical levels to the right. Additionally, the number of substructures for the city of London is almost twice that of Manhattan, which is another reason why the left plan is more living than the right one.)

Space has a healing effect, and this insight into space has been well established in the literature (e.g., Ulrich 1984). Human beings have an innate nature of loving lifelike things and processes such as forests and weathering. This affinity to nature is termed by the eminent biologist E. O. Wilson (1984) as biophilia. The biophilia effect has been used to help create living environments by integrating lifelike things such as light, water, and trees (Kellert et al. 2008). It should be noted that a true biophilia goes beyond the simple integration of natural things, but to create things that look like nature structurally (Salingaros 2015). Jackson Pollock (1912–1956) once said that he was not interested in mimicking nature, yet his poured paintings capture the order of nature. In this connection, living structure or the recurring notion of far more smalls than larges, as Alexander (2002–2005) has argued, appears to be the order that exists not only in nature, but also in what we make or build. The order – or living structure – constitutes the core or foundation of the new geography.

**6. The new geography, its implications, and future works**
The new geography laid down in the paper is established under the third view of space or on the solid foundation of living structure. The new geography is inclusive of a wide range of conventional disciplines, including for example architecture, urban design and planning, urban science, and regional science, all to do with how to transform our cities and communities to be more livable, more living or more beautiful. Thus, the new geography is a science of living structure, not only for understanding geographic forms and processes, but also – probably more importantly – for making and remaking geographic space to be living or more living (i.e., sustainable spatial planning or design). Table 3 lists the differences between conventional and new geography. The new geography goes beyond the two cultures under which science is separated from art (Snow 1959), towards the third culture (Brockman 1995) under which science and art is one. In the rest of this section, we further discuss on implications of the new geography and future works to be done.

Table 3: Comparison between the conventional and new geography

| Conventional geography | New geography |
|---|---|
| Mechanistic world view of Descartes | Organismic world view of Whitehead |
| First and second views of space of Newton and Leibniz | Third view of space of Alexander |
| Understanding geographic forms and processes | Understanding + making living structures |
| Tobler's law dominated | Scaling law dominated |
| A minor science or application of other major sciences | A major science or a science of living structure |

The new geography has huge implications on design and art, because goodness of art or design is no longer considered to be an arbitrary opinion or personal preference, but a matter of fact. It is essentially the underlying living structure that evokes a sense of goodness or beauty in the human mind and heart. Thus, there is a shared notion of quality or goodness of art among people or different peoples regardless of our cultures, faiths, and ethnicities. Goodness can be measured and quantified mathematically, and the outcome has over 70% agreement with people perception (e.g., Wu 2015, Salingaros and Sussman 2020). In this regard, the mirror-of-the-self experiment (Alexander 2002–2005) provides an effective measure for testing people on their judgement on goodness of things. In this experiment, two things or pictures (for example, Figures 1 and 6) are put side by side and human subjects are asked to provide their personal judgment to which one they have a higher degree of belonging or wholeness. The experiment is not kind of psychological or cognitive tests that seek inter-subjective agreement, but rather on degree of livingness, something objective or structural. This kind of experiment, as well as eye-tracking and other biometrics data (Sussman and Hollander 2015), constitutes an important future work in the new geography.

The new geography is a science of living structure, substantially based on living structure that resembles yet exceeds fractal geometry (Mandelbrot 1982). Like conventional geography, fractal geometry belongs to the camp of mechanistic thought. For example, the commonly used box-counting method for calculating fractal dimension is too mechanical, as the boxes defined at different levels of scale are not the right things (or the right perspective) for seeing living structure (cf., Section 3). As we have illustrated in Figures 1 and 2, we adopt an organismic rather than mechanistic way of seeing living structures. Fractals emerge from an iterative process, but the iterative process is often too strict or too exact. The real world is indeed evolved iteratively, but it is not as simple as fractals, neither classic fractals nor statistical fractals. Nature – naturally occurring things – has its own geometry, which is neither Euclidean nor fractal, but a living geometry that *"follows the rules, constraints, and contingent conditions that are inevitably encountered in the real world"* (Alexander 2002–2005). The major difference between fractal and living geometries lies probably on the two different world views. More importantly, goodness of a shape is not what fractal geometry concerned about, but it is the primary issue of living geometry.

Geographic information gathered through geographic information technologies has provided rich data sources for studying living structures on the Earth's surface from the perspectives of space, time, and human activities. This is particularly true for big data emerging from social media or the Internet. The big data are better than government owned or defined data for revealing the underlying living structure for two main reasons. First, big data have high resolution (like GPS locations of a couple of meters), and finer time scales (down to minutes and seconds for social media location data). Thus, they are better than government data for seeing living structure at different levels of scale. Second, government-defined spatial units, such as census tracts, are too rough or too arbitrary for seeing living structure. Instead, we should use naturally defined spatial units such as natural cities and auto-generated substructures (Jiang 2018, Jiang and Huang 2021), which are all defined from the bottom up, rather than imposed from the top down, thus making it easy to see living structures. While working with big data, we should try to avoid using grid-like approaches such as the digital elevation model. Although the digital elevation model has far more low elevations than high ones, the grid approach is not the right perspective for seeing living structures. Instead, we should use watersheds or water streams which are naturally or structurally defined. All these topics will be studied in the future for the new geography.

## 7. Conclusion

This paper is intended to help set geography on the firm foundation of living structure, based on the belief that how to make and remake livable spaces – or living structures in general – should remain at the core of geography. Considering a room, for example, we should first diagnose whether it is a living structure. If not, try to make it a living structure; if it is already, try to make it more living. This pursuit of living or more living structure extends from our rooms, gardens, buildings to streets, cities, and even the entire Earth's surface. Geography should not just be a minor science – as currently conceived under the Cartesian mechanistic world view – that seeks to apply other major sciences or technology for understanding geographic forms and processes (or city structure and dynamics in particular). This is because these major sciences have not yet solved the problem of how to do an effective making or creation. Instead, the problem of making or creating is commonly left to art, design, or engineering, where there is a lack of criteria for judging the quality or goodness of the created things. In this paper, geography is built on the criteria of living structure, not only for understanding geographic forms and processes, but also for transforming geographic space to be living or more living.

The new geography is founded on the third or organismic view of space, under which space is conceived as neither lifeless nor neutral, but a living structure capable of being more living or less living. The third view of space reveals that the nature of geographic space is a living structure or coherent whole, and its livingness or the degree of coherence can be quantified by the inherent hierarchy or the recurring notion of far more smalls than larges. Throughout this paper, we have attempted to argue that the scaling law should play a dominant role for it is universal, global, and across scales, whereas Tobler's law is available on each of these scales. These two laws are the two fundamental laws of living structure. To make a space living or more living, we must follow the two design principles or, more specifically, a series of biophilia design principles or the 15 structural properties. There are three fundamental issues about a geographic space (or a city in particular): (1) how it looks, (2) how it works, and (3) what it ought to be. The short response to these three issues is that a geographic space should look and work like a living structure and ought to become living or more living. Facing various challenges of our cities and environments, the new geography provides new concepts, questions, and solutions to tackle problems and to make and remake cities and communities to be more livable and more beautiful towards a sustainable society. It is time to transform conventional geography into the new geography, a science of living structure for the Earth's surface.

**Acknowledgement:**
I would like to thank the anonymous referees and the editor A-Xing Zhu for their constructive comments. In addition, Yichun Xie, Jia Lu, and Ge Lin read an earlier version of this paper, and Chris de Rijke helped with part of the figures. Thanks to you all. This project is partially supported by the Swedish Research Council FORMAS through the ALEXANDER project with grant number 2017-00824.

**Note 1 (on the state of the art of geography):**
To apply Alexander's remarkable insights on architecture, current state of the art of geography may be described by the following passage, in which I only changed architecture and architects to geography and geographers. All other words (Italic) are credited to Alexander, extracted from the preface he wrote for his biography by Grabow (1983), who spent six months with Alexander and tape-recorded over a hundred hours of conversations about his design thoughts. *In the past century,* geography *has always been a minor science – if it has been a science at all. Present day* geographers *who want to be scientific, try to incorporate the idea of physics, psychology, anthropology in their work, ... in the hope of keeping in tune with the "scientific" times. I believe we are on the threshold of a new era, when this relation between* geography *and the physical sciences may be reversed – when the proper understanding of the deep question of space, as they are embodied in* geography *will play a revolutionary role in the way we see the world and will do for the world view of the 21st and 22nd centuries, what physics did for the 19th and 20th.* The deep question touches the very nature of space or the organismic view of space, as formulated by Alexander, that space is neither lifeless nor neutral but a living structure capable of being more living or less living. It is initially the deep question or the organismic view of space that triggered me to develop this paper.